\documentclass[12pt]{article}
\pdfoutput=1
\usepackage{geometry,enumerate,amsmath,amssymb}
\usepackage{fullpage}
\usepackage{graphicx}
\usepackage{bm}
\numberwithin{equation}{section}
\newcommand{\be}{\begin{equation}}
\newcommand{\ee}{\end{equation}}
\newcommand{\bea}{\begin{eqnarray}}
\newcommand{\eea}{\end{eqnarray}}

\renewcommand{\hat}{\widehat}

\renewcommand{\epsilon}{\varepsilon}

\newcommand{\rt}{\sqrt{2}}
\newcommand{\eep}{e^{i\pi/4}}
\newcommand{\eem}{e^{-i\pi/4}}

\begin{document}
\title{
Spectral curves of hyperbolic monopoles from ADHM
}
\author{
  Paul Sutcliffe\\[10pt]
 {\em \normalsize Department of Mathematical Sciences,}\\
 {\em \normalsize Durham University, Durham DH1 3LE, United Kingdom.}\\ 
{\normalsize Email:  p.m.sutcliffe@durham.ac.uk}
}
\date{February 2021}

\maketitle
\begin{abstract}
  Magnetic monopoles in hyperbolic space are in correspondence with certain algebraic curves in mini-twistor space, known as spectral curves, which are in turn in correspondence with rational maps between Riemann spheres. Hyperbolic monopoles correspond to circle-invariant Yang-Mills instantons, with an identification of the monopole and instanton numbers, providing the curvature of hyperbolic space is tuned to a value specified by the asymptotic magnitude of the Higgs field. In previous work, constraints on ADHM instanton data have been identified that provide a non-canonical realization of the circle symmetry that preserves the standard action of rotations in the ball model of hyperbolic space. Here formulae are presented for the spectral curve and the rational map of a hyperbolic monopole in terms of its constrained ADHM matrix. This extends earlier results that apply only to the subclass of instantons of JNR type. The formulae are applied to obtain new explicit examples of spectral curves that are beyond the JNR class.
  \end{abstract}

\newpage
\section{Introduction}\quad
Magnetic monopoles in three-dimensional hyperbolic space share many of the features of their counterparts in Euclidean space \cite{At}. In both systems, there is a moduli space of solutions of the Bogomolny equation, with dimension $4N-1$ for monopoles of topological charge $N.$ Twistor methods provide a correspondence between points in the monopole moduli space and spectral curves, which are certain algebraic curves in the complex surface that is the space of oriented geodesics of three-dimensional space, whether it be hyperbolic or Euclidean.
If the curvature of hyperbolic space is equal to minus four times the square of the asymptotic magnitude of the Higgs field then hyperbolic monopoles are more tractable than their Euclidean cousins. In this tuned case, a hyperbolic monopole of charge $N$ is equivalent to a circle-invariant charge $N$ Yang-Mills instanton in four-dimensional Euclidean space, with the Higgs field of the monopole given by the component of the instanton gauge field associated with the circle action \cite{At}. This is the case of interest in the present paper, so this relation shall be assumed from now on, in units in which the tuned curvature is -1.

The ADHM construction \cite{ADHM} is a twistor transform mapping a Yang-Mills instanton to a quaternionic matrix satisfying a nonlinear reality condition, with the size of the matrix related to the charge of the instanton. A general study of the imposition of circle invariance within the ADHM construction was performed by Braam and Austin \cite{BA} and in the tuned case of interest here this simply correponds to a reduction of the quaternionic ADHM matrix to a complex matrix. This provides a transform between a hyperbolic monopole and a complex ADHM matrix. Furthermore, a simple formula exists for the spectral curve in terms of this matrix \cite{MuSi2,BCS}. It might therefore appear that the construction of spectral curves for hyperbolic monopoles is a solved problem in the tuned case. However, the difficulty with implementing this approach lies in finding complex ADHM matrices, which is further complicated by the fact that the natural circle action used by Braam and Austin yields the upper half space model of hyperbolic space, where spatial rotations act in a non-canonical manner. This has prevented the use of symmetry methods, despite the fact that symmetry reductions have been successfully applied in the past to find quaternionic ADHM matrices.

Some progress has been made \cite{MS} by restricting the Yang-Mills instanton to the subclass of JNR instantons \cite{JNR} given by a harmonic ansatz \cite{CF}. Invariance under the canonical circle action is simple to impose within this restriction and yields explicit results for all monopoles with $N=1,2,3.$ For $N>3$ this provides the construction of a $(3N+2)$-dimensional subspace of the $(4N-1)$-dimensional moduli space, in terms of free data given by $N+1$ points on the sphere together with a set of positive relative weights. Furthermore, an explicit expression for the spectral curve is also available in terms of this free data \cite{BCS}. However, for tuned hyperbolic monopoles that are beyond this JNR class, there is only one known example of a spectral curve \cite{NoRo}: this has $N=4$ and cubic symmetry and is obtained by working directly with the spectral curve and applying methods from algebraic geometry.

Imposition of a non-canonical circle action on Yang-Mills instantons leads to hyperbolic monopoles within the ball model of hyperbolic space, with the advantage that rotations are given by the standard action of rotations of the ball. Constraints on quaternionic ADHM matrices have been obtained \cite{MS} that realize this non-canonical circle action and are compatible with the symmetry methods used to construct examples of ADHM matrices beyond the JNR class. However, the spectral curves of the associated hyperbolic monopoles were not known. Here a simple formula is presented for the spectral curve of the monopole in terms of its
constrained quaternionic ADHM matrix. The previously known example of the $N=4$ spectral curve with cubic symmetry is reproduced within this formalism together with new examples that are beyond the JNR class.

Another characterization of a charge $N$ hyperbolic monopole is via a based degree $N$ rational map between Riemann spheres (modulo a phase) \cite{At2}, this being the hyperbolic analogue of Donaldson's rational map for Euclidean monopoles \cite{Do}. Braam and Austin \cite{BA} provided a formula for the rational map from its complex ADHM matrix and within the JNR class this provides a simple expression for the rational map in terms of the free data of weighted points on a sphere \cite{BCS}. Here a formula is presented for the rational map from its constrained quaternionic ADHM matrix.

\section{A tale of two circles}\quad
In standard form, the ADHM matrix for an $SU(2)$ charge $N$ Yang-Mills instanton in $\mathbb{R}^4$ is an $(N+1)\times N$ quaternionic matrix $\hat M.$
It is convenient to split this matrix into its top row, given by the non-zero vector $L$, and the remaining symmetric square part $M$, 
\be
\hat M = \begin{pmatrix} L\\ M \end{pmatrix}.
\label{Mhat}
\ee
The ADHM matrix is required to satisfy the reality condition that
$\hat M^\dagger \, \hat M$ is a real invertible $N\times N$ matrix, where
$^\dagger$ denotes quaternionic conjugate transpose.

Using the quaternionic representation of a point in $\mathbb{R}^4$,
$x=x_4 + x_1 i+ x_2 j + x_3 k,$ 
the conformal group of $\mathbb{R}^4$ acts as quaternionic M\"obius transformations
\be
x\mapsto x'=(Ax+B)(Cx+D)^{-1} \,.
\label{genct}
\ee
The canonical circle action used by Braam and Austin \cite{BA} is
\be
\begin{pmatrix} A&B\\ C & D \end{pmatrix}
=\begin{pmatrix}
e^{i\theta/2} & 0\\
0 & e^{-i\theta/2}  \end{pmatrix}.
\label{circle1}
\ee
The fixed point set of this circle action is the plane $x_4=x_1=0.$
The quotient of $\mathbb{R}^4$ with this plane removed is
conformal to hyperbolic space, naturally identified with the upper half space model
\be
ds^2=\frac{dx_2^2+dx_3^2+dr^2}{r^2},
\ee
with $r=\sqrt{x_1^2+x_4^2}>0.$ The requirement that the instanton is invariant under this circle action, and therefore descends to a hyperbolic monopole, is simply the restriction that the quaternionic ADHM matrix $\hat M$ is now complex. Lower case letters will be used to denote this restriction, so that a complex ADHM matrix is written as
\be
\hat m = \begin{pmatrix} \ell\\ m \end{pmatrix},
\label{complexadhm}
\ee
where $\hat m^\dagger \hat m$ is real, and $^\dagger$ is now the complex conjugate transpose.

The fixed point set of the non-canonical circle action
\be
\begin{pmatrix} A&B\\ C & D \end{pmatrix}=
\begin{pmatrix} \ \ \cos({\theta}/{2})&\sin({\theta}/{2})\\ -\sin({\theta}/{2}) & \cos({\theta}/{2})
\end{pmatrix},
\label{circle2}
\ee
is the unit sphere $x_1^2+x_2^2+x_3^2=1$ in the hyperplane $x_4=0.$ 
The quotient of $\mathbb{R}^4$ with this sphere removed is
conformal to hyperbolic space, naturally identified with the unit ball model
\be
ds^2=\frac{4(dX_1^2+dX_2^2+dX_3^2)}{(1-R^2)^2},
\ee
with $R=\sqrt{X_1^2+X_2^2+X_3^2}<1.$ The relation to 
the quaternionic representation of a point in $\mathbb{R}^4$ is
\be
x=\frac{(1-R^2)\sin\theta+2(X_1i+X_2j+X_3k)}{1+R^2+(1-R^2)\cos\theta},
\ee
with the fixed point set being $R=1.$
The advantage of this non-canonical circle action
is that it commutes with the standard rotations of the ball
\be
\begin{pmatrix} A&B\\ C & D \end{pmatrix}
=\begin{pmatrix} q&0\\ 0 & q \end{pmatrix},
\label{rotate}
\ee
where $q$ is a unit quaternion corresponding to the $SU(2)$ representative of the rotation.

The instanton is invariant under the circle action (\ref{circle2}), and therefore corresponds to a hyperbolic monopole, if its ADHM matrix satisfies the following three constraints \cite{MS}
\bea
&& M \mbox{\ is pure quaternion,}\label{con1}\\
&& \hat M^\dagger \, \hat M \mbox{\ is the identity matrix,}\label{con2}\\
&& LM=\mu L \,, \mbox{\ where $\mu$ is a pure quaternion.}\label{con3} 
\eea

It is perhaps worthwhile making a comment regarding the relation to instantons on $S^4.$ In terms of the geometry of $S^4,$ all circle actions are equivalent with a fixed point set $S^2.$ The quotient of $S^4$ with this $S^2$ removed is conformal to hyperbolic space. To go from instantons in $\mathbb{R}^4$ to instantons on $S^4$ requires compactification by the addition of a point at infinity. The canonical circle action has the point at infinity inside the fixed point set of the circle action, whereas the non-canonical circle action has the point at infinity outside the fixed point set. This distinguishes the two circle actions when considering instantons in $\mathbb{R}^4.$

At this stage, the relation between the two alternative ADHM descriptions of a hyperbolic monopole, as a quaternionic matrix satisfying the above three constraints, or as a complex matrix (\ref{complexadhm}), is not at all clear. At the end of this section a formula will be presented to obtain the second description from the first, but before this the formulae for the spectral curve and rational map of the monopole will be described in both manifestations.

The mini-twistor space of a real 3-manifold with constant curvature is the complex surface given by the space of its oriented geodesics \cite{Hi}. The mini-twistor space of three-dimensional hyperbolic space is (almost) $\mathbb{CP}^1 \times\mathbb{CP}^1,$
with coordinates $(\eta,\zeta)$ that are a pair of points on the
 Riemann sphere that is the boundary of the unit ball
 model of hyperbolic space. The associated oriented geodesic starts at 
 $-1/\bar\eta$ and ends at $\zeta$, so the anti-diagonal $\bar\eta\zeta=-1$ must be removed, hence the word almost in the previous sentence. 

 The spectral curve of a hyperbolic monopole is an algebraic curve in mini-twistor space satisfying certain reality and non-singularity conditions.
 In detail, the spectral curve of a charge $N$ hyperbolic monopole 
is a biholomorphic curve in $\mathbb{CP}^1 \times\mathbb{CP}^1$ 
of bidegree $(N,N).$
Writing the spectral curve as
\be \sum_{i=0,j=0}^N c_{ij}\eta^i\zeta^j=0, 
\label{gensc}
\ee
the reality condition on the complex constants $c_{ij}$, 
that follows from reversing the orientation of the geodesic,
is
\be
\bar c_{ij}=(-1)^{N+i+j}c_{N-j,N-i}.
\label{reality}
\ee
The spectral curve describes all geodesics along which
a certain linear operator, constructed from the monopole fields, has a normalizable solution. This is equivalent to imposing non-singularity conditions
\cite{At,MuSi1,MNS}
on the algebraic curve that can be written in terms of relations between integrals of holomorphic differentials around particular cycles. In general this is an intractable problem in algebraic geometry, however for low values of $N$ and in symmetric cases this line of attack is a viable, although complicated, method to obtain the spectral curve. The spectral curve of an $N=3$ monopole with tetrahedral symmetry and an $N=4$ monopole with cubic symmetry have been
obtained using this approach \cite{NoRo}.

An alternative procedure to obtain the spectral curve of a hyperbolic monopole uses its ADHM matrix. In terms of the complex ADHM matrix (\ref{complexadhm}) the spectral curve is \cite{MuSi2,BCS}    
\be
\mbox{det}\big(\eta\zeta m^\dagger+\zeta-\eta\hat m^\dagger \hat m-m\big)=0.
\label{complexdet}
\ee
The problem of imposing the non-singularity constraint on the spectral curve has now been transferred to the task of finding solutions of the ADHM condition that $\hat m^\dagger \hat m$ is real. Having simply shifted the difficulty, it is no surprise that the general $4N-1$ parameter family of solutions (up to equivalence) of this nonlinear equation cannot be found explicitly for arbitrary $N$. The JNR subfamily referred to earlier, that gives the general solution for $N\le 3$ and a $3N+2$ parameter family of solutions for $N>3$, corresponds to a conformal extension of the 't Hooft solution in which $m$ is diagonal and $\ell$ is real, so that $\hat m^\dagger \hat m$ is automatically real.

The rational map of a charge $N$ hyperbolic monopole is a degree $N$ based rational map between Riemann spheres, ${\cal R}(z)$, with the base point condition ${\cal R}(\infty)=0.$ A pair of rational maps are equivalent if they differ only by the multiplication by a constant phase.  There is a one-to-one correspondence between these equivalence classes of rational maps and the $(4N-1)$-dimensional moduli space of hyperbolic monopoles \cite{At,At2}. The rational map may be viewed as scattering data, in the background of the monopole, along the geodesic associated with the point $(\eta,\zeta)=(0,z)$ in mini-twistor space. 
The rational map corresponding to the complex ADHM matrix (\ref{complexadhm}) is given by \cite{BA}
\be {\cal R}(z)=\ell(z-m)^{-1}\ell^t.
\label{complexrat}
\ee

Some quaternionic solutions of the ADHM equation beyond the JNR family have been obtained by applying symmetry methods to reduce the difficulty by imposing finite subgroups of the $SU(2)$ group of spatial rotations (\ref{rotate}). However, as these rotations do not commute with the circle action (\ref{circle1}) then this approach is not easy to implement directly for complex ADHM matrices, and no solutions have yet been found for complex matrices beyond the JNR class. In contrast, the circle action (\ref{circle2}) commutes with these spatial rotations and this allows the construction of ADHM matrices that are beyond the JNR class and also satisfy the constraints, (\ref{con1}), (\ref{con2}), (\ref{con3}), required to represent hyperbolic monopoles. Unfortunately, as analogues of the formulae (\ref{complexdet}) and (\ref{complexrat}) were unknown for constrained ADHM matrices, the spectral curves and rational maps of the associated monopoles could not be calculated. Here this issue is resolved, by introducing the formulae required to calculate the spectral curve and the rational map from the constrained ADHM matrix.

Given a constrained ADHM matrix satisfying (\ref{con1}), (\ref{con2}), (\ref{con3}), extract a triplet of real $N\times N$ matrices $M_1,M_2,M_3$ from $M$ by writing $M=iM_1+jM_2+kM_3.$ The formula for the spectral curve is the simple expression
\be
\mbox{det}\big(\eta\zeta(M_1-iM_2)+\zeta(1-M_3)-\eta(1+M_3)-(M_1+iM_2)\big)=0.
\label{sc}
\ee
To compute the rational map is a little more complicated. 
Define the Hermitian matrix
\be
H=
(1-M_3)^{-1/2}\bigg(1+M_3-(M_1-iM_2)(1-M_3)^{-1}(M_1+iM_2)\bigg)(1-M_3)^{-1/2},
\label{H}
\ee
where $(1-M_3)^{-1/2}$ is the inverse of the principal square root of the matrix $1-M_3.$ In all the examples presented in this paper, as can be verified by direct calculation, $1-M_3$ is a positive definite matrix and $H$ has rank one.
The expectation is that the constraints (\ref{con2}) and (\ref{con3}) imply these conditions in general, although a proof has not yet been found.
Let $v$ be the unit-length eigenvector of $H$ corresponding to the non-zero eigenvalue $\lambda.$ Note that $v$ is defined up to multiplication by an overall phase. The rational map is given by
\be
   {\cal R}(z)=\lambda v^\dagger\bigg(z-(1-M_3)^{-1/2}(M_1+iM_2)(1-M_3)^{-1/2}\bigg)^{-1}\bar v.
   \label{rat}
   \ee
   Note the unexpected feature that the formulae (\ref{sc}), (\ref{H}), (\ref{rat}) involve only the square part of the ADHM matrix and not the vector $L.$ This implies that it must be possible to recover $L$ given $M.$ This is indeed the case and the explicit expression is given by
   \be
   L=\frac{
   (1-M_1^2-M_2^2-M_3^2)_\bullet
   + i[M_2,M_3]_\bullet
   + j[M_3,M_1]_\bullet
   + k[M_1,M_2]_\bullet}
   {\sqrt{(1-M_1^2-M_2^2-M_3^2)_{\bullet\bullet}}},
   \label{L}
     \ee
     where $(1-M_1^2-M_2^2-M_3^2)_\bullet$ denotes the first non-zero row of the matrix $1-M_1^2-M_2^2-M_3^2$ and the operation $P_\bullet$ on any other matrix $P$ denotes the corresponding row of $P.$ The double application $P_{\bullet\bullet}$ denotes the first non-zero entry of the row vector $P_\bullet.$ 
The properties required of the triplet $(M_1,M_2,M_3)$ so that the constraints (\ref{con2}) and (\ref{con3}) are satisfied by (\ref{L}) are not clear. 
   
To make contact with the complex form of the ADHM matrix the formula is
   \be
   \hat m
   = \begin{pmatrix} \sqrt{\lambda}v^\dagger \\  (1-M_3)^{-1/2}(M_1+iM_2)(1-M_3)^{-1/2}\end{pmatrix}.
   \label{complex}
   \ee
 Using (\ref{complex}) and the relation $\lambda vv^\dagger=H$, together with the definition (\ref{H}), gives
\be
\hat m^\dagger \hat m=(1-M_3)^{-1/2}(1+M_3)(1-M_3)^{-1/2},
\ee
which is obviously real, hence the ADHM condition is indeed satisfied.
It is simple to check that substituting (\ref{complex}) into (\ref{complexdet}) and (\ref{complexrat}) yields (\ref{sc}) and (\ref{rat}) respectively, thereby validating these formulae.

In Section \ref{sec:JNR} a couple of low charge JNR examples are presented to see how these known examples fit within the new formalism. Section \ref{sec:bJNR} contains some new spectral curves that are obtained by turning attention to constrained ADHM matrices that are not within the JNR class.

\section{Examples of JNR type}\label{sec:JNR}\quad
The $N=1$ hyperbolic monopole with position ${\bf X}=(X_1,X_2,X_3)$ inside the unit ball has an ADHM matrix that splits into the real triplet
\be
M_1=\frac{2X_1}{1+|{\bf X}|^2},\quad
M_2=\frac{2X_2}{1+|{\bf X}|^2},\quad
M_3=\frac{2X_3}{1+|{\bf X}|^2}.
\label{charge1}
\ee
The constraints (\ref{con2}) and (\ref{con3}) are satisfied by taking
$L=(1-|{\bf X}|^2)/(1+|{\bf X}|^2),$ as given by (\ref{L}).
Substituting (\ref{charge1}) into (\ref{sc}) gives the spectral curve 
\be
2\eta\zeta(X_1-iX_2)+\zeta(1+|{\bf X}|^2-2X_3)-\eta(1+|{\bf X}|^2+2X_3)-2(X_1+iX_2)=0,
\label{star}
\ee
corresponding to all geodesics through the point ${\bf X}.$
This is known as the star at ${\bf X}.$

The Hermitian matrix (\ref{H}) is just a number in this case, given by
\be
H=\bigg(\frac{1-|{\bf X}|^2}{1+|{\bf X}|^2-2X_3}\bigg)^2,
\ee
so the eigenvalue is $\lambda=H$ with the one-dimensional unit-length
eigenvector $v=1.$
Using these expressions, the formula (\ref{rat}) for the rational map yields
\be
   {\cal R}(z)=\frac{(1-|{\bf X}|^2)^2/(1+|{\bf X}|^2-2X_3)^2}
   {z-2(X_1+iX_2)/(1+|{\bf X}|^2-2X_3)},
\ee     
where the position of the pole taken together with the square root of the residue can be recognized as the upper half space coordinates of the point in hyperbolic space with ball coordinates ${\bf X}.$

The associated complex ADHM matrix obtained from (\ref{complex}) simply consists of these upper half space coordinates
\be
\hat m=\frac{1}{1+|{\bf X}|^2-2X_3}\begin{pmatrix}1-|{\bf X}|^2 \\ 2X_1+2iX_2 \end{pmatrix},
\ee
and the reality condition is clearly satisfied as
\be \hat m^\dagger \hat m=\frac{1+|{\bf X}|^2+2X_3}{1+|{\bf X}|^2-2X_3}.\ee

Turning to a slightly more complicated example,
the triplet of real matrices from the constrained ADHM matrix of an $N=3$ monopole with tetrahedral symmetry are \cite{MS,Hou}
\be
M_1=\frac{1}{\sqrt{3}}\begin{pmatrix}
  0 & 0 & -1 \\
  0 & 0 & 0  \\
  -1 & 0 & 0
\end{pmatrix},\quad
M_2=\frac{1}{\sqrt{3}}\begin{pmatrix}
  0 & 0 & 0 \\
  0 & 0 & 1  \\
  0 & 1 & 0
\end{pmatrix},\quad
M_3=\frac{1}{\sqrt{3}}\begin{pmatrix}
  0 & 1 & 0 \\
  1 & 0 & 0  \\
  0 & 0 & 0
\end{pmatrix}.
\ee
Applying (\ref{L}) gives the row vector $L=\frac{1}{\sqrt{3}}(1,-k,-i)$ that can be used to check that the constraints are indeed satisfied.
Substituting these matrices into the formula (\ref{sc}) reproduces the spectral curve
\be
(\eta-\zeta)^3+\frac{i}{\sqrt{3}}(\eta+\zeta)(\eta^2\zeta^2-1)=0,
\label{tet3}
\ee
first obtained using the direct algebraic geometric approach \cite{NoRo} and later via the JNR class, by taking the free data of four points on the sphere to be located at the vertices of a regular tetrahedron \cite{BCS}.   
The tetrahedral symmetry of this curve is manifest by its invariance under the
generators of the tetrahedral group
\be
(\eta,\zeta)\mapsto(-\eta,-\zeta),\qquad \qquad
(\eta,\zeta)\mapsto\bigg(\frac{\eta-i}{\eta+i},\frac{\zeta-i}{\zeta+i}\bigg).
\label{tetrotations}
\ee
In this example, the rank one Hermitian matrix (\ref{H}) is
\be
H=\frac{3}{2}\begin{pmatrix}
  1 & e^{i\beta} & 0\\
  e^{-i\beta} & 1 & 0 \\
  0 & 0 & 0 \end{pmatrix},
\ee
where the phase $e^{i\beta}=(5+i\sqrt{2})/(3\sqrt{3})$ has been defined for notational convenience. The non-zero eigenvalue of this matrix and the associated
unit-length eigenvector are
\be
\lambda=3, \qquad v=\frac{1}{\sqrt{2}}\begin{pmatrix}e^{i\beta} \\ 1 \\ 0\end{pmatrix},
  \ee
producing, via (\ref{rat}), the rational map 
\be {\cal R}(z)=\frac{5iz^2+\sqrt{3}}{\sqrt{3}z^3+iz}e^{-i(\beta+\pi/2)}. \ee
This agrees, up to the arbitrary constant phase, with the expression for
the rational map obtained from the associated JNR data \cite{BCS}. 

Using (\ref{complex}) provides the realization of this $N=3$ tetrahedral monopole as a complex ADHM matrix
\be
\hat m=\frac{\sqrt{3}}{2}\begin{pmatrix}
  \sqrt{2}e^{-i\beta} & \sqrt{2} & 0 \\
  0 & 0& s_--s_+\\
  0 & 0 &  s_-+s_+\\
  s_--s_+ & s_-+s_+ & 0
\end{pmatrix}
\ee
where $s_\pm=(i\pm 1)3^{-3/4}(\sqrt{3}\pm 1)^{-1/2}.$
By making use of the identity $s_-\bar s_+=i\sin\beta,$ it is 
easy to verify that $\hat m^\dagger \hat m$ is indeed real and is given by
\be
\hat m^\dagger \hat m
=\begin{pmatrix} 2 & \sqrt{3} & 0 \\
\sqrt{3} & 2 & 0\\
0 & 0 & 1
\end{pmatrix}.
\ee

\section{New spectral curves}\label{sec:bJNR}\quad
This section contains new examples of spectral curves that are obtained by using known ADHM matrices that satisfy the constraints (\ref{con1}), (\ref{con2}), (\ref{con3}), but are outside the JNR class.

 There is a one-parameter family, $\theta\in(-\pi/2,\pi/2)$, of tetrahedrally symmetric $N=4$ hyperbolic monopoles \cite{MS,LM}. Set
 $a=\frac{1}{\sqrt{3}}\sin\theta$ and $b=\frac{1}{2\sqrt{2}}\cos\theta$, then the real triplet of matrices is given by
    \be
   M_1=\begin{pmatrix}
   a & 0 & -b & -b \\
   0 & a & -b & b\\
   -b & -b & -a & 0 \\
   -b & b & 0 & -a
   \end{pmatrix},\quad
   M_2=\begin{pmatrix}
   a & -b & 0 & -b \\
   -b & -a & b & 0\\
   0 & b & a & -b \\
   -b & 0 & -b & -a
   \end{pmatrix},\quad
    M_3=\begin{pmatrix}
   a & -b & -b & 0 \\
   -b & -a & 0 & -b\\
   -b & 0 & -a & b \\
   0 & -b & b & a
   \end{pmatrix},
    \ee
 with $L=\frac{1}{2}\cos\theta\,(1,i,j,k)$, as obtained from (\ref{L}), confirming that the constraints are satisfied.  
   
 Applying (\ref{sc}) gives the spectral curve  
 \be
(\eta-\zeta)^4+\frac{8-5\cos^2\theta}{8+\cos^2\theta}(\eta^4\zeta^4+6\eta^2\zeta^2+4\eta\zeta(\eta^2+\zeta^2)+1)
 -\frac{16\sqrt{3}i\sin\theta}{8+\cos^2\theta}(\eta^2-\zeta^2)(\eta^2\zeta^2-1)=0,
 \label{tet4}
 \ee
 that is invariant under the tetrahedral generators (\ref{tetrotations}).
 
In the limit $\theta\to\pm \pi/2$ the curve becomes
\be
(\eta^4+1)(\zeta^4+1)+12\eta^2\zeta^2\mp 2i\sqrt{3}(\eta^2-\zeta^2)(\eta^2\zeta^2-1)=0,
\ee
which is the product of four stars (\ref{star}) with positions given by the vertices of a regular tetrahedron on the sphere at infinity in hyperbolic space, ${\bf X}=\frac{1}{\sqrt{3}}(\pm1,\pm1,\pm1)$ with an odd (even) number of minus signs for $\theta$ negative (positive). This agrees with the description of this family as four monopoles approaching the origin from infinity on the vertices of a tetrahedron and receding to infinity on the vertices of the dual tetrahedron, as $\theta$ varies through its allowed interval $(-\pi/2,\pi/2).$ At the midpoint, $\theta=0$, the symmetry is enhanced to cubic symmetry, with the additional generator $(\eta,\zeta)\to(i\eta,i\zeta)$, and this is the one member of this family of spectral curves that was already known \cite{NoRo},
\be
(\eta-\zeta)^4+\frac{1}{3}(\eta^4\zeta^4+6\eta^2\zeta^2+4\eta\zeta(\eta^2+\zeta^2)+1)=0.
\ee
Although this spectral curve for the $N=4$ monopole with cubic symmetry was already known, the associated rational map was not. However, it can now be obtained by using the expression (\ref{rat}). Restricting to the cubic case, the Hermitian matrix (\ref{H}) is
\be
H=\frac{2}{3}\begin{pmatrix}
  1 & -\eem & -\eep & -i \\
  -\eep & 1 & i & -\eem \\
  -\eem & -i & 1 & \eep \\
  i & -\eep & \eem & 1
\end{pmatrix},
\ee
with non-zero eigenvalue and associated eigenvector
\be
\lambda=\frac{8}{3}, \qquad\qquad
v=\frac{1}{2}\begin{pmatrix}-i \\ -\eem \\ \eep \\ 1 \end{pmatrix}.
\ee
Substituting these expressions into (\ref{rat}) gives the rational map
\be
   {\cal R}(z)=\frac{8\sqrt{3}z^2}{3z^4+1},
   \ee
   where the square symmetry is evident as the relation ${\cal R}(iz)=-{\cal R}(z).$
   
   Applying (\ref{complex}) yields the complex form of the ADHM matrix for the cubic $N=4$ monopole
   
       \resizebox{0.97\linewidth}{!}{%
         $\displaystyle
         \medmuskip=1mu
   \hat m=\frac{\sqrt{2}}{12}\begin{pmatrix}
     i4\sqrt{3} & -4\sqrt{3}\eep & 4\sqrt{3}\eem & 4\sqrt{3} \\

     2\eep+(\sqrt{3}-2)\rt\eem &
     \rt-4i &
     -4-\rt i &
     -(\sqrt{3}+2)\rt\eep-2 \eem \\

     \rt-4i &
     (\sqrt{3}-2)\rt\eep-2\eem &
     2 \eep-(\sqrt{3}+2)\rt\eem &
     4+\rt i \\

     -4-\rt i &
     2 \eep-(\sqrt{3}+2)\rt\eem &
     -(\sqrt{3}-2)\rt\eep+2 \eem &
     \rt -4i \\

     -(\sqrt{3}+2)\rt\eep-2 \eem &
     4+\rt i &
     \rt -4i &
     -2\eep -(\sqrt{3}-2)\rt\eem\\
          
   \end{pmatrix}
   $}
       \begingroup
       \setlength{\belowdisplayskip}{0pt}\setlength{\belowdisplayshortskip}{0pt} \be \ee
providing the first example of a complex ADHM matrix that is not within the JNR class.
\endgroup
The reality condition can be verified explicitly, 
   \be
   \hat m^\dagger \hat m =\begin{pmatrix}
     5 & -2\sqrt{2} & -2\sqrt{2} & 0\\
     -2\sqrt{2} & 5 & 0 & -2\sqrt{2} \\
     -2\sqrt{2} & 0 & 5 & 2\sqrt{2} \\
     0 & -2\sqrt{2} & 2\sqrt{2} & 5
   \end{pmatrix}.
   \ee
   
   There is an icosahedrally symmetric $N=7$ hyperbolic monopole
   that is related to the dodecahedron \cite{MS,SiSu}
   and has the triplet of real matrices
\be
M_1=\frac{1}{4}
\left( \begin {array}{ccccccc} 0&0&0&0&2&0&0\\ \noalign{\medskip}0&0&0
&0&0&0&0\\ \noalign{\medskip}0&0&0&0&0&0&\sqrt {5}+1
\\ \noalign{\medskip}0&0&0&0&0&\sqrt {5}-1&0\\ \noalign{\medskip}2&0&0
&0&0&0&0\\ \noalign{\medskip}0&0&0&\sqrt {5}-1&0&0&0
\\ \noalign{\medskip}0&0&\sqrt {5}+1&0&0&0&0\end {array} \right),
\ee
\be
M_2=\frac{1}{4}
\left( \begin {array}{ccccccc} 0&0&0&0&0&2&0\\ \noalign{\medskip}0&0&0
&0&0&0&\sqrt {5}-1\\ \noalign{\medskip}0&0&0&0&0&0&0
\\ \noalign{\medskip}0&0&0&0&\sqrt {5}+1&0&0\\ \noalign{\medskip}0&0&0
&\sqrt {5}+1&0&0&0\\ \noalign{\medskip}2&0&0&0&0&0&0
\\ \noalign{\medskip}0&\sqrt {5}-1&0&0&0&0&0\end {array} \right),
\ee
\be
M_3=\frac{1}{4}
\left( \begin {array}{ccccccc} 0&0&0&0&0&0&2\\ \noalign{\medskip}0&0&0
&0&0&\sqrt {5}+1&0\\ \noalign{\medskip}0&0&0&0&\sqrt {5}-1&0&0
\\ \noalign{\medskip}0&0&0&0&0&0&0\\ \noalign{\medskip}0&0&\sqrt {5}-1
&0&0&0&0\\ \noalign{\medskip}0&\sqrt {5}+1&0&0&0&0&0
\\ \noalign{\medskip}2&0&0&0&0&0&0\end {array} \right),
\ee
with (\ref{L}) producing the companion quaternionic vector
$
L=\frac{1}{2}(1,i,j,k,0,0,0).
$
Applying (\ref{sc}) produces the icosahedrally symmetric spectral curve
\be
\begin{split}
&\left( \eta-\zeta \right) ^{7}+\frac{\sqrt {5}}{60} \left(\eta -\zeta
\right)\times \\
& \left(  ( 3+\sqrt {5} )  \left( \eta+\zeta
 \right) ^{2}-2 \left( \eta\,\zeta-1 \right) ^{2} \right)  \left( 
 ( 3-\sqrt {5} )  \left( \eta+\zeta \right) ^{2}+2
 \left( \eta\,\zeta+1 \right) ^{2} \right)
 \left(  ( \sqrt {5}\,
 \eta\,\zeta-1) ^{2}+4 \right)\\
 &=0.
 \end{split}
\ee
The order two symmetry, $(\eta,\zeta)\mapsto(-\eta,-\zeta),$ of this spectral curve is clear but the remaining symmetries of the icosahedral group, of order three and five, are not as transparent. They are given by
\be
\medmuskip=1mu
(\eta,\zeta)\mapsto \bigg(
\frac{(2-i(\sqrt{5}+1))\eta+1-\sqrt{5}}{(\sqrt{5}-1)\eta+2+i(\sqrt{5}+1)},
\frac{(2-i(\sqrt{5}+1))\zeta+1-\sqrt{5}}{(\sqrt{5}-1)\zeta+2+i(\sqrt{5}+1)}
\bigg),
\label{icos_c3}
\ee
and
\be
\medmuskip=1mu
(\eta,\zeta)\mapsto \bigg(
\frac{(2i-\sqrt{5}-1)\eta+i(\sqrt{5}-1)}{i(\sqrt{5}-1)\eta-2i-\sqrt{5}-1},
\frac{(2i-\sqrt{5}-1)\zeta+i(\sqrt{5}-1)}{i(\sqrt{5}-1)\zeta-2i-\sqrt{5}-1}
\bigg),
\ee
respectively. These formulae are simply the standard formulae in a different orientation, as follows. A more convenient form of the spectral curve can be obtained by applying a rotation so that the order five symmetry becomes a rotation around a Cartesian axis. After applying the rotation
\be
(\eta,\zeta)\mapsto \bigg(
\frac{(\sqrt{5}-1)\eta-(2+\sqrt{10-2\sqrt{5}})}{-\eta(2+\sqrt{10-2\sqrt{5}})+\sqrt{5}-1},
\frac{(\sqrt{5}-1)\zeta-(2+\sqrt{10-2\sqrt{5}})}{-\zeta(2+\sqrt{10-2\sqrt{5}})+\sqrt{5}-1}
\bigg),
\ee
followed by the rotation $(\eta,\zeta)\mapsto(e^{i\pi/5}\eta,e^{i\pi/5}\zeta)$,
the spectral curve takes the more compact form
 \be
 \left( \eta-\zeta \right)  \big( \eta \left( {\eta}^{5}-2 \right) 
 \left( 2{\zeta}^{5}+1 \right) +\zeta\left( {\zeta}^{5}-2
 \right)  \left( 2{\eta}^{5}+1 \right) +25{\eta}^{2}{\zeta}^{2}
 \left( {\eta}^{2}+{\zeta}^{2} \right)  \big)=0. 
\ee
In this orientation the order five symmetry is 
\be
(\eta,\zeta)\mapsto (\omega\eta,\omega\zeta),
\label{orderfive}
\ee
where $\omega=e^{2\pi i/5}$, and the order three symmetry is
\be
(\eta,\zeta)\mapsto \bigg(
\frac{(\omega^3-1)\eta+\omega-\omega^2}{(\omega-\omega^2)\eta+1-\omega^3},
\frac{(\omega^3-1)\zeta+\omega-\omega^2}{(\omega-\omega^2)\zeta+1-\omega^3}
\bigg).
\label{orderthree}
\ee
As a final example, there is an $N=17$ hyperbolic monopole with icosahedral symmetry \cite{MS,Su17} that is related to the truncated icosahedron and has a real triplet of matrices that can be written in the block matrix form
\be
M_i=\frac{1}{24}\begin{pmatrix}
0 & 4T_i^t & 0 & 4U_i^t \\
4T_i & 0 & \varepsilon_{ijk}T_jU^t_k & 0   \\
0 & \varepsilon_{ijk}U_kT^t_j & 0 & 2\varepsilon_{ijk}U_jU^t_k \\
4U_i & 0 & 2\varepsilon_{ijk}U_kU^t_j & 0
\end{pmatrix}.
\ee
Here $\varepsilon_{ijk}$ is the totally antisymmetric tensor and the matrices that appear in the blocks are given by
\be
T_1=
\left( \begin {array}{ccc} 2&0&0\\ \noalign{\medskip}0&0&0
\\ \noalign{\medskip}0&0&\sqrt {5}+1\\ \noalign{\medskip}0&\sqrt {5}-1
&0\end {array} \right),\quad
U_1=
\left( \begin {array}{ccc} \sqrt {2}&0&0\\ \noalign{\medskip}0&0&0
\\ \noalign{\medskip}-\sqrt {2}\sqrt {3}&0&0\\ \noalign{\medskip}0&-
\sqrt {5}-1&0\\ \noalign{\medskip}0&0&\sqrt {5}-1\end {array} \right),
\ee
\be
T_2=
\left( \begin {array}{ccc} 0&2&0\\ \noalign{\medskip}0&0&\sqrt {5}-1
\\ \noalign{\medskip}0&0&0\\ \noalign{\medskip}\sqrt {5}+1&0&0
\end {array} \right),\quad
U_2=
\left( \begin {array}{ccc} 0&\sqrt {2}&0\\ \noalign{\medskip}0&0&
\sqrt {5}+1\\ \noalign{\medskip}0&\sqrt {2}\sqrt {3}&0
\\ \noalign{\medskip}\sqrt {5}-1&0&0\\ \noalign{\medskip}0&0&0
\end {array} \right),
\ee
\be
T_3=
\left( \begin {array}{ccc} 0&0&2\\ \noalign{\medskip}0&\sqrt {5}+1&0
\\ \noalign{\medskip}\sqrt {5}-1&0&0\\ \noalign{\medskip}0&0&0
\end {array} \right),\quad
U_3=
\left( \begin {array}{ccc} 0&0&-2\,\sqrt {2}\\ \noalign{\medskip}0&-
\sqrt {5}+1&0\\ \noalign{\medskip}0&0&0\\ \noalign{\medskip}0&0&0
\\ \noalign{\medskip}-\sqrt {5}-1&0&0\end {array} \right).
\ee
The companion quaternionic vector (\ref{L}) is
$ L=\frac{1}{2}(0,0,0,1,i,j,k,0,0,0,0,0,0,0,0,0,0).$

Using (\ref{sc}) and applying the same rotations as in the previous example, to
obtain a convenient orientation, the spectral curve is
\begin{align}
  \begin{split}
(\eta-\zeta)\bigg(&
\left( -3{\eta}^{15}+41{\eta}^{10}-33{\eta}^{5}+1 \right) {
\zeta}^{16}+ \left( -3{\eta}^{16}+3{\eta}^{11}-51{\eta}^{6}-11
\eta \right) {\zeta}^{15}\\&
+ \left( 438{\eta}^{12}+768{\eta}^{7}+162
{\eta}^{2} \right) {\zeta}^{14}+ \left( 338{\eta}^{13}+468{\eta}
^{8}-338{\eta}^{3} \right) {\zeta}^{13}\\&
+ \left( 438{\eta}^{14}-132
{\eta}^{9}+1462{\eta}^{4} \right) {\zeta}^{12}+ \left( 3{\eta}^{
  15}+5025{\eta}^{10}-6291{\eta}^{5}+33 \right) {\zeta}^{11}\\&
+ \left( 41{\eta}^{16}+5025{\eta}^{11}+4273{\eta}^{6}+51\eta
 \right) {\zeta}^{10}+ \left( -132{\eta}^{12}-17252{\eta}^{7}-768
       {\eta}^{2} \right) {\zeta}^{9}\\&
       + \left( 468{\eta}^{13}+648{\eta}^
{8}-468{\eta}^{3} \right) {\zeta}^{8}+ \left( 768{\eta}^{14}-17252
{\eta}^{9}+132{\eta}^{4} \right) {\zeta}^{7}\\&
+ \left( -51{\eta}^{
15}+4273{\eta}^{10}-5025{\eta}^{5}+41 \right) {\zeta}^{6}+ \left( 
-33{\eta}^{16}-6291{\eta}^{11}-5025{\eta}^{6}+3\eta \right) {
  \zeta}^{5}\\&
+ \left( 1462{\eta}^{12}+132{\eta}^{7}+438{\eta}^{2}
 \right) {\zeta}^{4}+ \left( -338{\eta}^{13}-468{\eta}^{8}+338{
   \eta}^{3} \right) {\zeta}^{3}\\&
 + \left( 162{\eta}^{14}-768{\eta}^{9}
+438{\eta}^{4} \right) {\zeta}^{2}+ \left( -11{\eta}^{15}+51{
  \eta}^{10}+3{\eta}^{5}+3 \right) \zeta\\&
+{\eta}^{16}+33{\eta}^{11}+
41{\eta}^{6}+3\eta
\bigg)=0.
\end{split}
\end{align}
This icosahedrally symmetric curve is invariant under the generators
(\ref{orderfive}) and (\ref{orderthree}) of the icosahedral group.
It would seem to be a challenge to obtain this spectral curve directly using
the approach of algebraic geometry rather than the ADHM method used here.

\section{Conclusion}\quad
A simple formula has been presented for the spectral curves of hyperbolic monopoles in terms of ADHM matrices satisfying conditions that imply circle invariance of the associated instantons and commuting rotations that act canonically on the ball model of hyperbolic space. This is a hyperbolic analogue of the construction of spectral curves from Nahm data for monopoles in Euclidean space \cite{Nahm,Hit}. In both contexts this provides an integrable systems approach to obtain the spectral curve that is more tractable than the direct methods of algebraic geometry.

The results in this paper are applicable when the curvature of hyperbolic space is tuned so that the charge of the hyperbolic monopole is equal to the charge of the associated circle-invariant instanton. A similar relation between hyperbolic monopoles and instantons exists for an infinite set of discrete values of the curvature, with the instanton charge being any integer multiple of the monopole charge \cite{At}. For the canonical circle action (\ref{circle1}) this yields a discrete integrable system for complex matrices on a lattice, with the number of lattice points equal to the ratio of the instanton and monopole charges \cite{BA}. A generalization of the expression (\ref{complexdet}) provides the formula for the spectral curve \cite{MuSi2}. However, for the circle action (\ref{circle2}) the generalization of the constraints (\ref{con1}), (\ref{con2}), (\ref{con3}) on ADHM matrices is an open problem once the instanton charge is greater than the monopole charge. It would be interesting to find the appropriate generalization of the constraints and to see if a formula similar to (\ref{sc}) provides the spectral curve in this more general situation.

\section*{Acknowledgements}
Many thanks to Nick Manton for useful discussions.

\end{document}